\documentclass[aps, pre, twocolumn, amsmath, superscriptaddress,showkeys,showpacs]{revtex4-1}

\usepackage{commath}
\usepackage[innercaption]{sidecap}
\usepackage{amssymb}
\usepackage{graphicx}
\usepackage{graphicx,epstopdf}
\usepackage{xcolor, soul}
\usepackage{dcolumn}
\usepackage{bm}
\usepackage{mathrsfs} 
\usepackage{amsmath} 
\usepackage[colorlinks=true,linkcolor=blue,citecolor=blue]{hyperref}%
\usepackage[many]{tcolorbox}
\usepackage{fontenc}
\usepackage{float}
\usepackage{amsthm}
\usepackage{subfigure}
\usepackage{color}
\usepackage{ragged2e}
\usepackage{enumerate}
\usepackage{hyperref}
\def\be{\begin{equation}}
\def\ee{\end{equation}}
\def\bea{\begin{eqnarray}}
\def\eea{\end{eqnarray}}
\def\bfg{\begin{figure}[H]}
\def\efg{\end{figure}}

\begin{document}

\title{Mortality makes coexistence vulnerable in evolutionary game of rock-paper-scissors}

\author{Sirshendu Bhattacharyya} 

\email{sirs.bh@gmail.com}

\affiliation{Department of Physics, R.R.R Mahavidyalaya,
Radhanagar, Hooghly 712406, India}

\author{Pritam Sinha}  
\affiliation{Physics and Applied Mathematics Unit, Indian Statistical Institute, Kolkata 700108, India}

\author{Rina De}
\affiliation{Department of Physics, R.R.R Mahavidyalaya,
	Radhanagar, Hooghly 712406, India}

\author{Chittaranjan Hens}

\email{chittaranjanhens@gmail.com}

\affiliation{Physics and Applied Mathematics Unit, Indian Statistical Institute, Kolkata 700108, India}

\begin{abstract}
\noindent
Multiple species in the ecosystem are believed to compete cyclically for survival and thus maintain balance in nature. Stochasticity has also an inevitable role in this dynamics. Considering these attributes of nature, the stochastic dynamics of the rock-paper-scissor model based on the idea of cyclic dominance  becomes an effective tool to capture different aspects of ecosystem. The evolutionary dynamics of this model crucially depends on different interactions representing different natural habits. In this framework we explore the role of mortality of individual organism in the collective survival of a species. For this purpose a new parameter called `natural death' is introduced. It is meant for bringing about the decease of an individual irrespective of any intra- and interspecific interaction. We perform Monte Carlo simulation followed by the stability analysis of different fixed points of defined rate equations and observe that the natural death rate is surprisingly one of the most significant factors in deciding whether an ecosystem would come up with a coexistence or a single species survival.
\end{abstract}

\maketitle


\noindent
Earth's ecosystem consists of a diverse population where the constituent species continuously strive to keep in existence. The mechanism of existence of multiple competing species has been a long studied issue in ecology \citep{smith1982evolution,nowak2006evolutionary}. In physics, theoretical explorations mainly from the perspective of evolutionary game theory, have been plunged into different directions to understand this natural process \citep{may1972will,hauert2005game,szabo2007evolutionary,roca2009evolutionary}.  Numerous models have been proposed for the same purpose and one of those models is the Rock-Paper-Scissor (RPS) model which, under different formalism (Lotka-Volterra and May-Leonard) \citep{lotka1920analytical,volterra1926variazioni,leonard1975nonlinear}, have been widely studied. As the name suggests, the model in its simplest form, demonstrates that biodiversity can be maintained through interspecific cyclic competition which gives all the species a fair chance to survive \cite{kerr2002local,szolnoki2014cyclic, shi2010basins,he2010spatial,reichenbach2008self,avelino2014interfaces,park2017emergence,roman2013interplay, lutz2013intransitivity, vukov2013diverging, szabo2008self,cheng2014mesoscopic}. The interspecific competition is mainly described by predator-prey type interaction. However the model has been studied incorporating other types of interactions as well (see \citep{szolnoki2014cyclic} for review). Most of the studies have mainly investigated the cyclic competition in order to find coexistence \citep{schreiber2013spatial,cheng2014mesoscopic,laird2014population,park2017emergence}. 
Biological examples have also been presented in this regard. For instance, the morph prevalence of three-morph mating system in the side-blotched lizard \cite{sinervo1996rock} can be captured by this model. The cyclic dominance of Pacific salmon \cite{guill2011three} or stable state in microbes \cite{kerr2002local} are some other examples.

A notable feature is that the coexistence is generally associated with cyclic dominance and oscillations in the dynamics of species densities and small changes of system parameters are proved to matter a lot for this state. It has been observed that variables like mobility, intensity of environmental noise etc. can be a determining factor behind the rapid extinction of species \cite{reichenbach2008instability, reichenbach2007mobility,reichenbach2007noise}. Another recent study shows that in presence of death by starvation, the stability of the coexisting species can only be sustained if the reproduction rate is significantly high \cite{avelino2019death}. These examples establish the fact that, the coexistence being vulnerable, various system parameters decide whether the ecosystem would support or jeopardize the stable biodiversity. Apart from this, certain interaction rates sometimes give rise to the abundance of the weakest species (with respect to hunting/predation) \cite{avelino2019predominance}. This case has been confirmed for both conservative (Lotka-Voltera formulation) and non-conservative (May-Leonard formulation) number of individuals in a large square lattice in the backdrop of non-spatial formulation. In addition, examples are also present where such parameters (e.g. mobility) may cause the revival of stability amongst the species through re-emergence of several spiral patterns in the dynamics of the system \cite{avelino2012junctions,jiang2011effects,wang2011pattern,bazeia2019invasion}.

As the existence of a species is highly dependent on the parameters denoting relative probabilities of interactions, situations may also arise that a non-uniform or asymmetric interaction rate in the stochastic process of cyclic competition lead to the extinction of a particular species \cite{berr2009zero}. This theoretical analysis is found to be consistent with the experiment of \emph{E.Coli}, in which one species (strain)  eventually survives in a well mixed population \cite{kerr2002local}. In addition to the asymmetric and species-dependent interaction rates, mobility is considered to be another important parameter to drive the system towards a single species dominated state \citep{shi2010basins,venkat2010mobility}.
 However, system size may appear to be a crucial factor in these cases \cite{menezes2019uneven}. Although we know that the stochastic fluctuations have surprising role on the dynamics of this type of systems, the fact that one out of three (or more) species outcompetes the other in presence of a cyclic dominance is still an intriguing issue. The question that in which circumstances a defined ecosystem is left with only one of its constituent species has been the motivation of our present work. 

\par In this article we explore the competition of survival among three species each having tendency of predation and reproduction under favorable condition. In addition we consider the fact that every organism has a finite lifetime. Hence every species should have a rate of death which is independent of its interaction with any other individual. We call it the rate of `natural death'. With this rate introduced, the dynamics with cyclic dominance is investigated in $2$-dimensional square lattice under May-Leonard formalism. We perform Monte Carlo simulation followed by analytical treatment deterministic differential equation to find out the effect of the death rate on the system. Our study finds the natural death to be an important parameter which controls the system's pathway more than the other two. As compared to the predation and reproduction rates, a small variation of death rate leads to a change of final state of the system \--- from coexistence to single species survival.



\begin{figure*}[ht!]
	\includegraphics[scale=0.5]{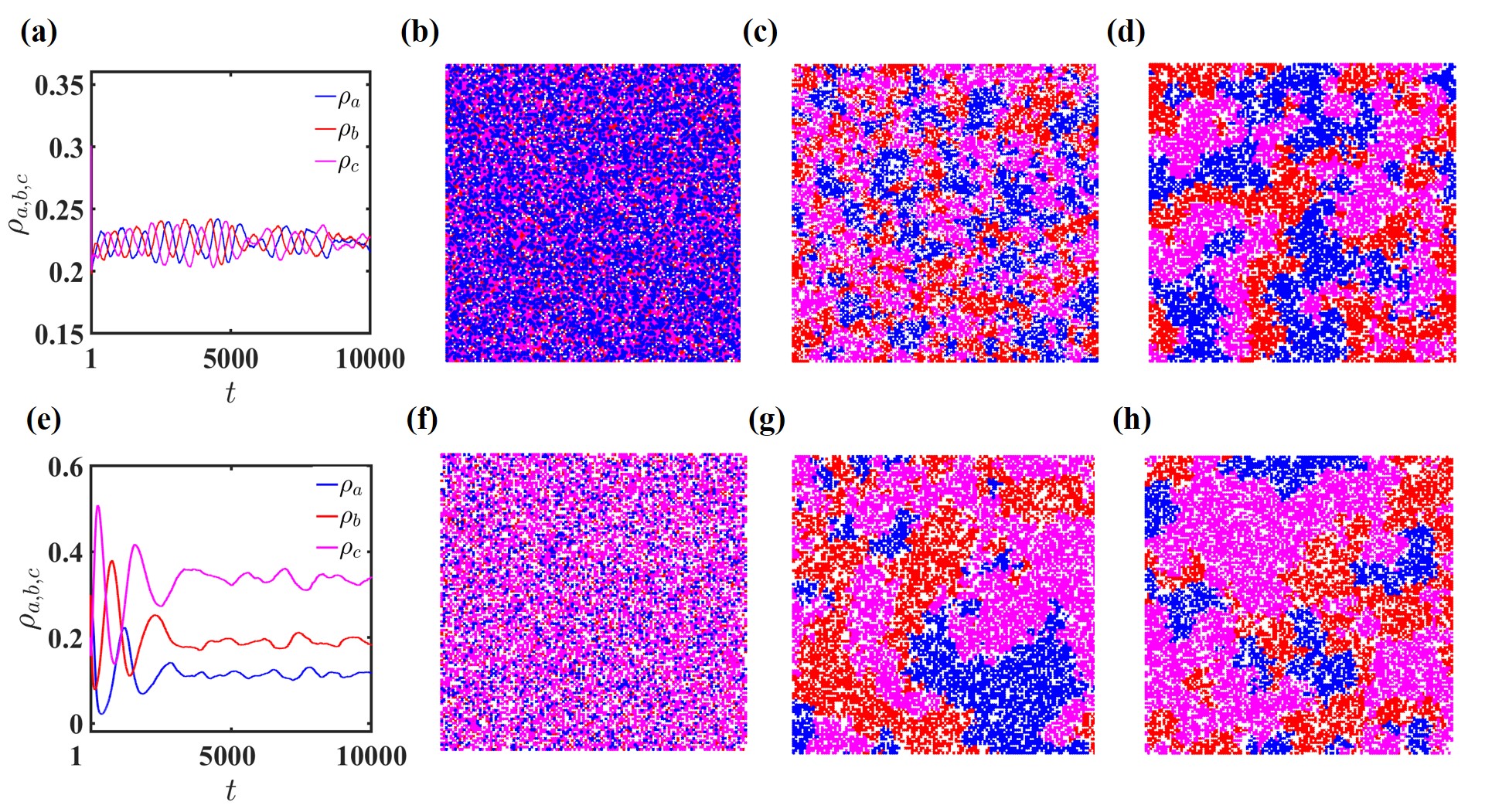}
	\caption{Dynamics of the densitis of three species studied on $500 \times 500$ lattice. (a) - (d): Spatio-temporal plot for $p_{a,b,c}=0.2$, $r_{a,b,c}=0.4$, $d_{a,b,c}=0.1$. (a) represents the temporal behavior of the densities and (b), (c) \& (d) show the spatial distribution of the species at $t=0,\;700\; \&\; 10^4$ respectively. (e) - (h): Spatio-temporal plots for $p_{a,b,c}=0.2$, $r_{a,b,c}=0.4$, $d_{a,c}=0.1$, and $d_b=0.12$. Here, the long time behavior of the densities indicates the predominance of species $C$ following a large-scale variation at initial times. The blue, red and magenta colors correspond to $A$, $B$ and $C$ respectively and the white portion in the spatial distribution denotes empty site. The spatial plots (g) - (h) are again captured at $t=0,\;700\; \&\; 10^4$.}
	\label{fig1}
\end{figure*}



\noindent
We consider a 3-species RPS model mapped in a 2-dimensional square lattice where each site of the lattice is supposed to either contain a single member of any of the three species or left vacant. Likewise the idea of children's rock-paper-scissor game, the three components (species) of our model undertake a cyclic interaction in terms of predation. In addition, the elements of the species have the prospect of reproduction or the risk of death at any time. Here the term death may be interpreted as natural or accidental death for which no other member of the system is responsible. Hence we call it natural death to differentiate it from the death caused by competition. The introduction of this parameter requires to adopt the May-Leonard formulation \cite{leonard1975nonlinear} where, unlike the Lotka-Volterra formulation, the total number of individuals is not conserved. Following this formulation we have also assumed that, in the process of  evolutionary game dynamics, the predation strategy can create vacant site in the adjacent neighbor and conversely the action of reproduction replaces a vacant site with an individual \citep{avelino2019predominance}. Therefore, if the normalized species abundance of $A, B$ and $C$ are $\rho_a$, $\rho_b$ 
and $\rho_c$ respectively, the underline conservation rule will be
$\rho_a+\rho_b+\rho_c=1-\rho_v$, 
where $\rho_v$ is the fraction of vacant site with respect to total number of sites.

In the cyclic process, we can write the predation strategy with following set of interactions
\bea
A + B & \longrightarrow & A + V \;\;\; \mbox{with rate}\;\; p_a\nonumber \\
B + C & \longrightarrow & B + V \;\;\; \mbox{with rate}\;\; p_b\nonumber \\
C + A & \longrightarrow & C + V \;\;\; \mbox{with rate}\;\; p_c
\label{predation}
\eea
where $V$ denotes a vacant site. Apart from the predation factor, each species may generate its own offspring, depending on its own reproduction probability and if the interacting site is vacant. The reproduction equations of each species can be captured by
\bea
A + V & \longrightarrow & A + A \;\;\; \mbox{with rate}\;\; r_a\nonumber \\
B + V & \longrightarrow & B + B \;\;\; \mbox{with rate}\;\; r_b\nonumber \\
C + V & \longrightarrow & C + C \;\;\; \mbox{with rate}\;\; r_c
\label{reproduction}
\eea
As mentioned earlier, we have additionally assumed that an individual from any species may die anytime with a specified probability, i.e.,
\bea
A + \varphi & \longrightarrow & V + \varphi \;\;\; \mbox{with rate}\;\; d_a \nonumber \\
B + \varphi & \longrightarrow & V + \varphi \;\;\; \mbox{with rate}\;\; d_b\nonumber \\
C + \varphi & \longrightarrow & V + \varphi \;\;\; \mbox{with rate}\;\; d_c
\label{death}
\eea
where $\varphi$ represents any of the three species or a vacancy. The natural death of an individual is thus implemented by transmuting it into an empty one.

We study the stochastic dynamics of the above model in a 2-dimensional 
$N\times N$ lattice with periodic boundary conditions applied. The dynamics of the Monte Carlo simulation is based on the interaction with any one of four non-diagonal nearest neighbors at a time. The simulation starts from a randomly chosen initial densities of three species ($\rho_{a}^{0}, \rho_{b}^{0}, \rho_{c}^{0}$) such that $\rho_{a}^{0}+ \rho_{b}^{0}+\rho_{c}^{0}= 1 - \rho_{v}^{0}$ where $\rho_{v}^{0}$ is the initial density of vacant sites. At each Monte Carlo step, a primary site is randomly chosen and, if it is not an empty site, one of its four nearest neighbors is again selected randomly. For a non-empty nearest neighbor, the two sites perform predation-prey interaction with probability, $p_{a,b,c}$ as the case may be. If the nearest neighbor is found to be empty, the fellow in the primary site attempts reproduction with probability, $r_{a,b,c}$. In addition to these two possible actions, according to  Eq.~(\ref{death}), the individual residing in the primary site may also abolish with a probability, $
d_{a,b,c}$ making the corresponding site vacant. The time unit of our calculation is defined by $N^2$ Monte Carlo steps.


We find a range of parameters where the system exhibits coexistence of all the constituent species. This kind of result has been reported earlier in ref.~\citep{avelino2019predominance}, where the effect of natural death was absent. Two different cases of coexistence have been presented through spatio-temporal plots in Fig.~\ref{fig1}. We have primarily taken 
$p_a = p_b = p_c=0.2$ 
in both the cases. The first row of the figure shows a case where we have taken $r_a = r_b = r_c=0.4$ and $d_a = d_b = d_c=0.1$ as well. All the three species having equal rate of predation, reproduction and death results in a coexistence with all the species densities oscillating around same value. Fig.~\ref{fig1}(a) shows that densities in this case oscillate with small amplitude around $0.22$. Figs.~\ref{fig1} (b-d) report the spatio-temporal pattern of the coexisting states. At $t=0$ (Fig.~\ref{fig1}(b)), the species are randomly distributed in a $500 \times 500$ lattice where blue, red, magenta and white colors represent the density of the species $A$, $B$, $C$ and the vacancy respectively. Another distribution is shown at $t\sim700$ unit (Fig.~\ref{fig1}(c)) where small scattered patches of each species can be observed in the spatial domain. After evolution over a large time, the size of the patches is increased (Fig.~\ref{fig1}(d)). The vacant sites (white) are seen to be embedded in these patches. This kind of colony-like pattern gets repeated over time with the patches changing their position and size with the total area for a particular color changing very little. In this kind of coexistence, no species can predominate over a long time.

However, we obtain a different situation by altering only $d_b$ to $0.12$. As shown in Fig.~\ref{fig1}(e)-(h), the coexistence in this case occurs along with the well built predominance of one particular species. The fluctuations are larger than that of the previous case. After an initial large-scale variation, the density of the species $C$ oscillates around $0.35$, whereas densities of $B$ and $A$ do the same around $0.2$ and $0.1$ respectively (Fig.~\ref{fig1}(e)). The predominance of species $C$ is also confirmed in spatial domain shown in Figs.~\ref{fig1}(g) and (h). 
The enhancement in the death rate of species $B$ alters the fate of other species: species $C$ becomes most abundant whereas the population of species $A$ becomes minimum. This behavior is cognizable from physical point of view. The increase of death rate of $B$ ensures lesser preying of $C$ and thereby indirectly favors $C$ to outcompete $A$. What makes the death rate more special is the susceptibility of the system's diversity to it. As compared to the other parameters, a small change in the death rate significantly changes the dynamics and the final state. In both the cases discussed above or even in the other cases of coexistence, we observe the densities always swinging about the mean value. This feature can be explained in terms of stability and we will show that this points where the constituents coexist is the unstable fixed point of the mean-field deterministic model. The fluctuations are notably high during coexistence and it is even higher in the case of asymmetrical value(s) of the parameters. This can be interpreted as the effect of environmental noise which becomes more effective when interaction rates lose any kind of symmetry.

\begin{figure}
\includegraphics[height=6.0cm, width=6.5cm]{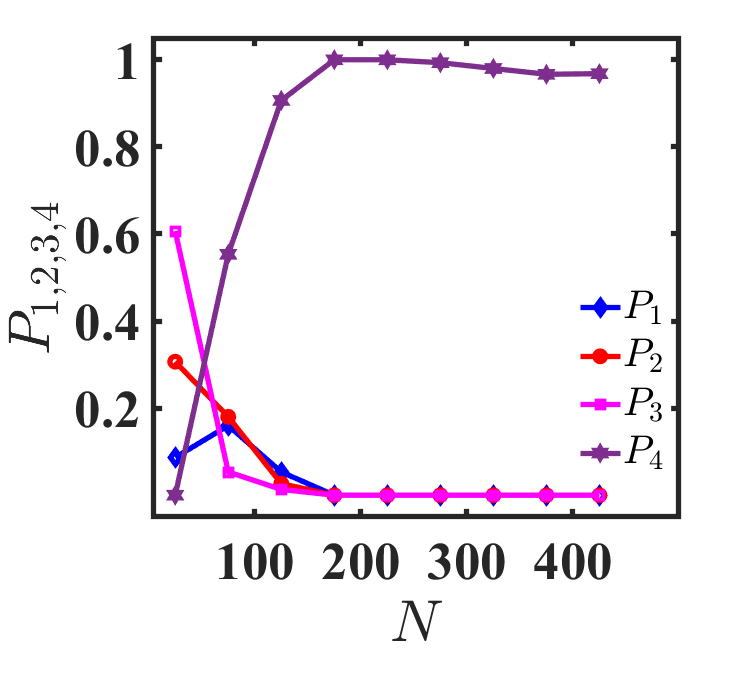}
\caption{Probabilities of reaching a final state varying with linear system size $N$. $P_{1,2,3}$ denotes the probability of the system reaching a state where only $A$, $B$ or $C$ survives respectively. $P_4$ denotes the probability of coexistence. For this calculation, the rates are taken to be $d_{a,b,c}=0.1$, $r_{a,b,c}=0.4$, $p_{a,b,c}=0.2$. The probabilities have been estimated from $2\times 10^2$ realizations.}
\label{fig2}
\end{figure}

We have also observed that these cases of coexistence occurs for large system size. However the final state is found to vary for small systems under the same parameter values. For very small system-size, for the same parameters as in Fig.~\ref{fig1}(a), the system has very low probability to end in a coexistence. Rather the survival of one solitary species ($A$, $B$ or $C$) is preferred there. As we go on increasing the lattice size, the dynamics gradually becomes inclined to go towards the coexistence. It can be understood from Fig.~\ref{fig2} where we have calculated the probability of the system going to a single species state or coexistence over a large number of initial configurations. For this purpose we have introduced the linear system size, $N$. The probabilities are seen to vary with $N$. However for 
$N \gtrapprox 200$, 
the probabilities become independent of $N$ and the coexistence becomes one and only destiny of the system.

\begin{figure}
\hspace*{-5mm}
\includegraphics[height=4.6cm, width=9.2cm]{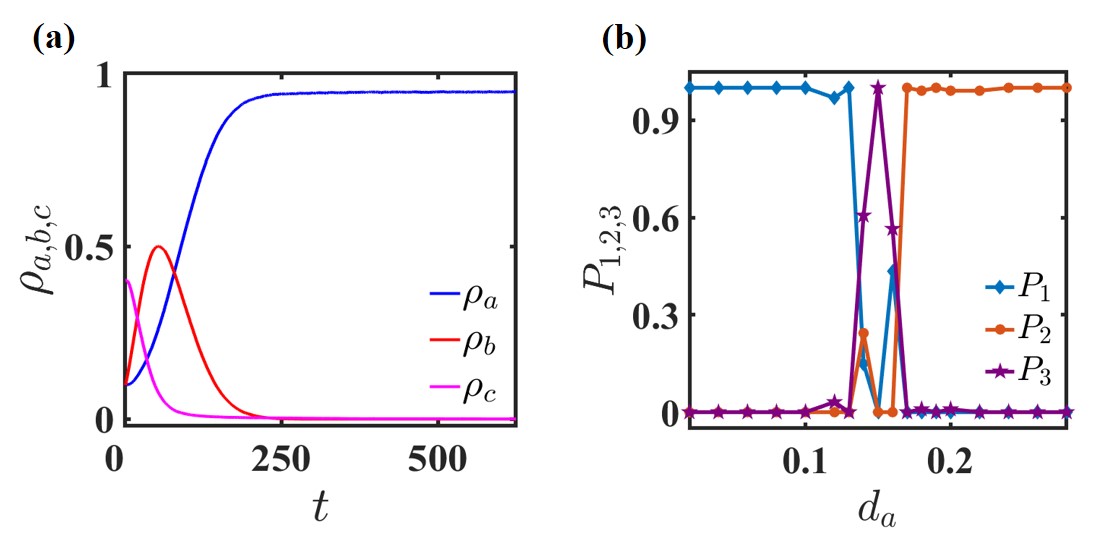}
\caption{(a) Dynamics of the densities of three species studied on $500 \times 500$ lattice for $p_{a,b,c}=0.2$, $r_{a,b,c}=0.4$, $d_{a}=0.02$, $d_b=d_c=0.15$. (b) Probabilities of final states with the variation of $d_a$ only. For lower values of $d_a$ the system prefers the survival of species $A$ only. $P_1$ denotes the probability of survival of $A$ only. On increasing $d_a$ the probability of coexistence (denoted by $P_3$) increases  and it becomes $1$ at $d_a=0.15\;(=d_{b,c})$. The survival of only species $B$ (probability denoted by $P_2$) eventually becomes preferable to the system when $d_a$ increases further.}
\label{fig3}
\end{figure}

A remarkable behavior of the system is observed in a different range of natural death rates. For some values of $d_{a,b,c}$ 
the dynamics leads to the survival of a single species. Fig.~\ref{fig3}(a) shows one of such cases where two species have equal death rate ($d_b=d_c=0.15$) 
while the third one having moderately lower rate of death ($d_a=0.02$). In this case the two species with higher death rate get abolished leaving only one species to survive in course of time. The oscillatory dynamics as seen in the case of coexistence cannot be observed here and the variation of system size cannot alter the final state as well. The final state is found to be robust also with respect to considerable changes of predation and reproduction rates.
The destiny of the system is however found to be dependent on the death rates. Fig.~\ref{fig3}(b) shows the system's probability of reaching a final state with varying death rate of one species (here we consider species $A$). When species $A$ has considerably lower death rate than that of the other two, the ecosystem favors the survival of species $A$ only although all the species have same rate of predation and reproduction. The same system prefers coexistence as the final state if $d_a$ becomes comparable to the other death rates. The probability of coexistence becomes $1$ at $d_a=d_b=d_c$. For larger $d_a$, again the case of single species survival prevails, but this time with species $B$ being alive. This case is non-trivial in the sense that the death rates of $B$ and $C$ are equal. We also observe that the absence of death rates drives the system towards the coexistence and the survival of only one species cannot be obtained by varying the rates of predation and reproduction.

\noindent
In our model, the abundance of a species is captured by two of its intrinsic parameters: reproduction rate which increases the number and natural death rate which is able to reduce it. In addition, the species abundance is controlled by the predation rate of its preadator. Inserting these ingredients into the rate equations, we can construct three coupled mean-field equations. Thus the temporal evolution of the densities of the three species can be described by these differential equations: 
\begin{eqnarray}
\dfrac{\partial \rho_a}{\partial t} &=& \rho_a(t) \left [ {r_{a}} \rho_v(t)- p_{c}\rho_c(t)-d_{a}\right] = f(\rho_a,\rho_b,\rho_c) \nonumber \\[0.2cm]
\dfrac{\partial  \rho_b}{\partial t} &=& \rho_b(t) \left [ r_{b} \rho_v(t)- p_{a} \rho_a(t)-d_{b}\right] = g(\rho_a,\rho_b,\rho_c) \nonumber \\[0.2cm]
\dfrac{\partial \rho_c}{\partial t} &=& \rho_c(t) \left [ r_{c} \rho_v(t)- p_{b} \rho_b(t)-d_{c}\right] = h(\rho_a,\rho_b,\rho_c)\nonumber
\label{rate-eq}
\end{eqnarray}
where $\rho_a(t)$, $\rho_b(t)$ and $\rho_c(t)$ 
are the densities of species 
$A$, $B$ and $C$ at time $t$ 
and $\rho_v(t) = 1- \left[\rho_a(t)+\rho_b(t)+\rho_c(t)\right]$ is the density of vacant sites. The system has a trivial (parameter independent) fixed point $FP1 \equiv ({\rho_a^{\ast}},\;\rho_b^{\ast},\;\rho_c^{\ast})\;=\;(0,\;0,\;0)$. The stability of the fixed point is checked through the \emph{Jacobian} ($J$) of the underlying dynamics. $J$ is constructed through the small perturbation of the flow around the fixed point and is captured by  
\begin{gather}
J =
\begin{pmatrix}
\dfrac{\partial f}{\partial \rho_a}\; & \;\dfrac{\partial f}{\partial \rho_b}\; & \;\dfrac{\partial f}{\partial \rho_c} \\[0.5cm]
\dfrac{\partial g}{\partial \rho_a}\; & \;\dfrac{\partial g}{\partial \rho_b}\; & \;\dfrac{\partial g}{\partial \rho_c} \\[0.5cm]
\dfrac{\partial h}{\partial \rho_a}\; & \;\dfrac{\partial h}{\partial \rho_b}\; & \;\dfrac{\partial h}{\partial \rho_c} \\
\end{pmatrix}_{(\rho_a^{\ast},\;\rho_b^{\ast},\;\rho_c^{\ast})}.
\label{jacmatrix} 
\end{gather}
The fixed point ($FP1$) will be stable when the death rate of each species will be grater than its own reproduction rate i.e, $d_a>r_a,\; d_b>r_b ~\mbox{and}~  d_c>r_c$  \cite{suppl-mat}. This is expected and trivial as the death rate of each species dominates in the competition here, resulting in extinction of all species.

There exists another set of fixed points where one species exists and the others are abolished. These fixed points look like:
 $FP21 \;\equiv\;\left(1-d_a/r_a,\;0,\;0\right)$,
 $FP22 \;\equiv\;\left(0,\;1-d_b/r_b,\;0\right)$ and 
 $FP23 \;\equiv\;\left(0,\;0,\;1-d_c/r_c\right)$.
 It is comprehensible that this set of fixed points corresponds to the single species survival (Fig.~\ref{fig3}(a)). The eigenvalue analysis of $J$ around $FP21$ reveals that it will be stable if 
$d_a<r_a$, $d_a \leq d_b$ and $d_a<d_c$ 
where we have considered $r_a=r_b=r_c$ and $p_a=p_b=p_c$ \cite{suppl-mat}. 
One may note that the relations of the stability condition holds well if we take $d_a=d_b$. This means that species $A$ may dominate over species $B$ even when they have identical predation, reproduction and death rates.
The entire set of conditions reveals that, in general, one species will be stable if its reproduction rate is considerably high and at the same time, death rate is less than that of its predator. These physical conditions are manifested in the stability of $FP22$ and $FP23$ as well \cite{suppl-mat}. 
The parameter sets chosen for the investigation of single species survival using  Monte Carlo simulation obey the conditions derived above. In Fig.~\ref{fig3}(a), for the chosen values of the parameters ($p_{a,b,c}=0.2$, $r_{a,b,c}=0.4$, $d_{a}=0.02$, $d_{b,c}=0.15$), the fixed point, $FP21:\;(0.95,0,0)$     is stable as the conditions are satisfied.
Now, keeping $d_b$ and $d_c$ fixed, if we increase $d_a$ from $0.02$ to $0.32$, there is transition in the survival of species (Fig.~\ref{fig3}(b)). Clearly, $FP21$ is stable  $d_a < d_{b,c}$. For large values of $d_a$, $FP22$ will be stable as $d_b < d_a$ and $d_b = d_c$ there. 
In the region between the two extremities i.e. around $d_a \approx 0.15$, there lies a competition between these fixed points because both set of conditions are favorable to be satisfied. Therefore, in this region we see the state of coexistence taking the opportunity. The fixed point representing the coexistence is another of its kind.
\par
This nontrivial fixed point is 
$FP3 \;\equiv\; (\rho_a^{\ast},\rho_b^{\ast},\rho_c^{\ast})\;=\;\left[ (r_b Q - d_b)/p_a,\; (r_c Q - d_c)/p_b,\; (r_a Q - d_a)/p_c) \right]$
where 
$Q = \left[ 1+(\frac{d_{a}}{p_{c}}+\frac{d_{b}}{p_{a}}+\frac{d_{c}}{p_{b}})\right]/\left[1+(\frac{r_{a}}{p_{c}}+\frac{r_{b}}{p_{a}}+\frac{r_{c}}{p_{b}})\right]$. The non-zero value of all the densities readily signifies the coexistence. Now, if we consider, 
$p_{a,b,c}=0.2$, $r_{a,b,c}=0.4$, $d_{a,b,c}=0.1$, we get, $FP3:\;
(0.22,0.22,0.22)$, which is unstable by nature \cite{suppl-mat}. The Monte Carlo simulation reveals that the densities of species in this case oscillate around this fixed point (Fig.~\ref{fig1}(a)). This is a special situation of the system when it has no available stable fixed point. Hence it has no other choice than to go towards the state of coexistence. However, the oscillatory dynamics around this fixed point is due to its unstable nature.
The asymmetric solution described in Fig.~\ref{fig1}(e-h) (predominance of species $C$ and reduced abundance of species $A$) also gives the same impact here.  
Therefore, specific settings of the parameters can reveal the coexistence in Monte Carlo simulation although they are not stable according to the deterministic model.
\par The sharp dependence of the system towards the change of death rate can be understood also from the rate equations. Let us denote the first, second and third term of each rate equation involving reproduction, predation and death respectively as $\mathcal{T}_r$ $\mathcal{T}_p$ and $\mathcal{T}_d$. Now, considering the density, $\rho_{a,b,c}\sim \mathcal{O}(1/N^2)$, 
one can show that $\mathcal{T}_{r,p}\sim \mathcal{O}(1/N^4)$ 
and $\mathcal{T}_d\sim \mathcal{O}(1/N^2)$. As a consequence the system's behavior principally depends on the death rate.

Therefore, both the simulation and the deterministic rate equation suggest that a suitable choice of death rate is enough to stabilize a fixed point giving non-zero density of one solitary species. Even a single death rate parameter can alter the existence of a species. The transition from one state to the other of completely different type by small variation of death rate (Fig.\ \ref{fig3}(b)) is quite interesting and is confirmed here by simulation and deterministic dynamics as well. 
However, near the transition point ($d_a = 0.15 \pm \epsilon$, for small $\epsilon$ at Fig.~\ref{fig3}(b)), the stability analysis gives no stable fixed points whereas the Monte Carlo simulation shows comparable probabilities of coexistence and different single species survival states. We would like to explore this region analytically in more details in future incorporating stochasticity in the rate equations.

\par
We study the dynamics of a three-component ecosystem mimicked by a $2$-dimensional square lattice with nearest neighbor interaction. Besides interspecific predation in cyclic manner, the species evolution is governed by reproduction and death as well. The Monte Carlo simulation following the above protocol reveals that the influence of death rate is quite significant in maintaining biodiversity. We show that specific death rates are able to exterminate the biodiversity and mark the survival of only one species in the ecosystem. Again, at the time of coexistence, they play an important role in determining the predominance of a species. We could substantiate our result with the help of stability analysis of the differential equation for this model. The reason of death rate being foremost controlling parameter could also be understood.
Our approach unveils possible reasons for asymmetric coexistence and single-species dominance in evolutionary game dynamics. In presence of natural or accidental death which is not instigated by intervention of other individual, the atypical vulnerability of the fate of the cyclically interacting species in an ecosystem is an interesting phenomenon and has not been explored before. We would like to investigate in future the impact of natural death in some extended RPS models mapped in a complex network. It would be also interesting to study the effect of noise and disorder in this framework. \\

\noindent
SB acknowledhges Supravat Sarkar for useful discussion.  CH is supported by the INSPIRE-Faculty grant (code: IFA17-PH193).

\bibliography{bibliography_maintext}
\bibliographystyle{apsrev4-1}

\onecolumngrid
\pagebreak
\begin{center}
\textbf{\Large Supplemental Material: Mortality makes coexistence vulnerable in evolutionary game of rock-paper-scissors}
\end{center}
\setcounter{equation}{0}
\setcounter{figure}{0}
\setcounter{table}{0}
\setcounter{page}{1}
\makeatletter
\renewcommand{\theequation}{S\arabic{equation}}
\renewcommand{\thefigure}{S\arabic{figure}}
\renewcommand{\bibnumfmt}[1]{[S#1]}
\renewcommand{\citenumfont}[1]{S#1}

\section*{Stability analysis of deterministic Rock-Paper-Scissor models}
\subsection{Model description}
\label{Model}
The dynamics of RPS model can be captured by
\begin{eqnarray}
\frac{\partial\rho_a}{\partial t} &=& \rho_a(t) \left [ r_{a}\rho_v(t)- p_{c}\rho_c(t)-d_{a}\right] \\
\frac{\partial\rho_b}{\partial t} &=& \rho_b(t) \left [ r_{b}\rho_v(t)- p_{a}\rho_a(t)-d_{b}\right] \\
\frac{\partial\rho_c}{\partial t} &=& \rho_c(t) \left [ r_{c}\rho_v(t)- p_{b}\rho_b(t)-d_{c}\right],
\label{Dynamics_1}   
\end{eqnarray}
with a constraint $ \rho_v = 1-\rho_a-\rho_b-\rho_c$.
\subsection{Equilibrium points}
The first trivial fixed point is 
\begin{equation}
FP1:\;\rho_a^{\ast}=0,\rho_b^{\ast}=0,\rho_c^{\ast}=0\;\;\&\;\rho_v^{\ast}=1
\label{fp1}
\end{equation}
The second set of fixed points where a single species exist:
\begin{eqnarray}
FP21:\; \rho_a^{\ast}=1-\frac{d_{a}}{r_{a}}, \rho_b^{\ast}=0,\rho_c^{\ast}=0 ~ \&~ \rho_v^{\ast}=d_{a}/r_{a}.\label{fp21}\\
FP22:\;  \rho_a^{\ast}=0, \rho_b^{\ast}=1-\frac{d_{b}}{r_{b}},\rho_c^{\ast}=0 ~ \&~  \rho_v^{\ast}=d_{b}/r_{b}. \label{fp22}\\
FP23:\; \rho_a^{\ast}=0, \rho_b^{\ast}=0,\rho_c^{\ast}=1-\frac{d_{c}}{r_{c}} ~ \&~   \rho_v^{\ast}=d_{c}/r_{c}. \label{fp23}
\end{eqnarray}
There is another fixed point where all the species have non-zero densities. We call it as $FP3$ which looks like
\begin{equation}
FP3:\;\rho_a^{\ast}=\frac {r_{b}Q-d_{b}}{p_{a}},\;
\rho_b^{\ast}=\frac {r_{c}Q-d_{c}}{p_{b}},\;
\rho_c^{\ast}=\frac {r_{a}Q-d_{a}}{p_{c}}
\label{fp3}
\end{equation}
where, $Q=\dfrac{1+(\frac{d_{a}}{p_{c}}+\frac{d_{b}}{p_{a}}+\frac{d_{c}}{p_{b}})}{1+(\frac{r_{a}}{p_{c}}+\frac{r_{b}}{p_{a}}+\frac{r_{c}}{p_{b}})}$

\subsection{Stability analysis of the equilibrium points}
We have to make small perturbation of the system around a fixed point to find the nature of stability. We can proceed further, by calculating the partial derivative of the flow of each variable. Let us assume
\begin{eqnarray}
\frac{\partial\rho_a}{\partial t}= \rho_a\left[ r_{a}(1-\rho_a-\rho_b-\rho_c)-p_{c}\rho_c-d_{a}\right] &=& f \nonumber \\
\frac{\partial\rho_b}{\partial t}=\rho_b\left[r_{b}(1-\rho_a-\rho_b-\rho_c)-p_{a}\rho_a-d_{b}\right]&=& g \nonumber \\
\frac{\partial\rho_c}{\partial t}=\rho_c\left[r_{c}(1-\rho_a-\rho_b-\rho_c)-p_{b}\rho_b-d_{c}\right] &=& h
\end{eqnarray}
The partial derivative of $f$ is as follows
\begin{eqnarray}   
\frac{\partial f}{\partial\rho_a}&=&-2r_{a}\rho_a-r_{b}\rho_b-r_{c}\rho_c-p_{c}\rho_c+r_{a}-d_{a},\\
\frac{\partial f}{\partial\rho_b}&=&-r_{a}\rho_a,\\
\frac{\partial f}{\partial\rho_c}&=&-r_{a}\rho_a-p_{c}\rho_a.
\end{eqnarray}
Now, considering the dynamical evaluation of the second species ($B$),  
the partial derivatives of $g$ are
\begin{eqnarray}
\frac{\partial g}{\partial\rho_a}&=&-r_{b}\rho_b-p_{a}\rho_b\\,
\frac{\partial g}{\partial\rho_b}&=&-2r_{b}\rho_b-r_{b}\rho_c-r_{b}\rho_a-p_{a}\rho_a+r_{b}-d_{b}\\
\frac{\partial g}{\partial\rho_ c}&=&-r_{b}\rho_b.
\end{eqnarray}
Similarly,
\begin{eqnarray}
\frac{\partial h}{\partial\rho_a}&=&-r_{c}\rho_c,\\
\frac{\partial h}{\partial\rho_b}&=&-r_{c}\rho_c-p_{b}\rho_c,\\
\frac{\partial h}{\partial\rho_c}&=&-2r_{c}\rho_c-r_{c}\rho_a-r_{c}\rho_b-p_{b}\rho_b+r_{c}-d_{c}.
\end{eqnarray}
Inserting all the terms in Jacobian matrix 
($J$)
 we can write 
\begin{gather}
J =
\begin{pmatrix}
 \frac{\partial f}{\partial\rho_a} & \frac{\partial f}{\partial\rho_b} & \frac{\partial f}{\partial\rho_c} \\[0.1cm]
 \frac{\partial g}{\partial\rho_a} & \frac{\partial g}{\partial\rho_b} & \frac{\partial g}{\partial\rho_c} \\[0.1cm]
 \frac{\partial h}{\partial\rho_a} & \frac{\partial h}{\partial\rho_b} & \frac{\partial h}{\partial\rho_c} \\
\end{pmatrix}_{(\rho_a^{\ast},\rho_b^{\ast},\rho_c^{\ast})}. 
\label{Jmatrix} 
\end{gather}
For a fixed point to be stable, the real part of all the eigen values must be negative.
\subsubsection{Stability analysis of $FP1$}
The first type of fixed point, $FP1$ is written in Eq.~(\ref{fp1}) for which we can write from  Eq.~(\ref{Jmatrix}),  
\begin{gather}
J =
\begin{pmatrix}
r_{a}-d_{a}&0&o\\
0&r_{b}-d_{b}&0\\
0&0&r_{c}-d_{c}\\
\end{pmatrix}
\label{stmatrix1} 
\end{gather}
Hence, for this fixed point to be stable
\begin{tcolorbox}
\begin{equation}
r_{a}<d_{a},\;
r_{b}<d_{b} \;\;\&\;\;
r_{c}<d_{c}.
\end{equation}
\end{tcolorbox}
\subsubsection{Stability analysis of $FP21$, $FP22$ and $FP23$}
We start with Eq.~(\ref{fp21}). 
The Jacobian in this case can be written as
\begin{gather}
J =
\begin{pmatrix}
-2r_{a}(1-\frac{d_{a}}{r_{a}})+r_{a}-d_{a}&-r_{a}(1-\frac{d_{a}}{r_{a}})&-(r_{a}+p_{c})(1-\frac{d_{a}}{r_{a}})\\
0&-(r_{b}+p_{a})(1-\frac{d_{a}}{r_{a}})+r_{b}-d_{b}&0\\
0&0&-r_{c}(1-\frac{d_{a}}{r_{a}})+r_{c}-d_{c}
\end{pmatrix}
\label{Fp21matrix} 
\end{gather}
The conditions for all the eigen values to be negative are:
\begin{tcolorbox}
\begin{equation}
d_{a}-r_{a}<0, ~~ \\
r_{b}\frac{d_{a}}{r_{a}}-d_{b}-p_{a}(1-\frac{d_{a}}{r_{a}})<0, ~~\&~~
r_{c}\frac{d_{a}}{r_{a}}-d_{c}<0.\\
\end{equation}
\end{tcolorbox}
\noindent
If $r_a=r_b=r_c$, we can rewrite the conditions as
\begin{equation}
d_a<r_a,\; d_a<d_c \;\;\rm{and}\;\; d_a\leq d_b
\end{equation}
It may be noted that the stability condition also holds for $d_a>d_b$ when $p_{a}(1-\frac{d_{a}}{r_{a}})>>0$. We have not explored this region in our work.\\
Now, from Eq.~(\ref{fp22}) 
we can again construct the Jacobian as
\begin{gather}
J =
\begin{pmatrix}
-r_{a}(1-\frac{d_{b}}{r_{b}})+r_{a}-d_{a}&0&0\\
-(r_{b}+p_{a})(1-\frac{d_{b}}{r_{b}})&-2r_{b}(1-\frac{d_{b}}{r_{b}})+r_{b}-d_{b}&-r_{b}(1-\frac{d_{b}}{r_{b}})\\
0&0&-(r_{c}+p_{b})(1-\frac{d_{b}}{r_{b}})+r_{c}-d_{c}.\\
\end{pmatrix}
\label{Fp22matrix} 
\end{gather}
and the stability conditions look like
\begin{tcolorbox}
\begin{equation}
r_{a}\frac{d_{b}}{r_{b}}-d_{a}<0,~~
d_{b}-r_{b}<0, ~ \& ~
r_{c}\frac{d_{b}}{r_{b}}-d_{c}-p_{b}(1-\frac{d_{b}}{r_{b}})<0.
\end{equation}
\end{tcolorbox}
\noindent
Imposing $r_a=r_b=r_c$ again simplifies the condition as
\begin{equation}
d_b<r_b,\; d_b<d_a, \;\;\rm{and} \;\; d_b\leq d_c
\end{equation}

Eq.~(\ref{fp23}) gives the third fixed point of this set and 
proceeding in similar way as in the previous two cases, the Jacobian here is given by
\begin{gather}
J =
\begin{pmatrix}
-(r_{a}+p_{a})(1-\frac{d{_c}}{r_{c}})+r_{a}-d_{a}&0&0\\
0&-r_{b}(1-\frac{d_{c}}{r_{c}})+r_{b}-d_{b}&0\\
-r_{c}(1-\frac{d_{c}}{r_{c}})&-(r_{c}+p_{c})(1-\frac{d_{c}}{r_{c}})&-2r_{c}(1-\frac{d_{c}}{r_{c}})+r_{c}-d_{c}.
\end{pmatrix}
\label{FP23matrix} 
\end{gather}
\noindent
producing the condition of stability
\begin{tcolorbox}
\begin{equation}
r_{a}\frac{d_{c}}{r_{c}}-d_{a}-p_{c}(1-\frac{d_{c}}{r_{c}})<0,~~
r_{b}\frac{d_{c}}{r_{c}}-d_{b}<0, ~ {\&}~
d_{c}-r_{c}<0.
\end{equation}
\end{tcolorbox}
\noindent
As discussed above, if $r_a=r_b=r_c$, we may write 
\begin{equation}
d_c<r_c,\; d_c<d_b \;\;\rm{and}\;\; d_c\leq d_a 
\end{equation}
\subsubsection{Stability analysis of $FP3$}
\noindent
Eq.~(\ref{fp3}) gives another set of fixed point which has three non-zero densities. The Jacobian matrix derived from this point is given by
\begin{equation}
 J =
\begin{pmatrix}
-r_{a}\rho_a^{\ast}&-r_{a}\rho_a&-r_{a}\rho_a^{\ast}-p_{c}\rho_a^{\ast}\\
-r_{b}\rho_b^{\ast}-p_{a}\rho_b^{\ast} & -r_{b}\rho_b^{\ast}&-r_{b}\rho_b^{\ast}\\
-r_{c}\rho_c^{\ast}&-r_{c}\rho_c^{\ast}-p_{b}\rho_c^{\ast}&-r_{c}\rho_c^{\ast}
\end{pmatrix}
\end{equation}
The characteristic equation for this matrix is
\begin{equation}
\lambda^{3}-Tr(J)\lambda^{2}+\left(C_{11}+C_{22}+C_{33}\right)\lambda-Det(J)=0
\end{equation}
where $C_{ii}$'s ($i=1,2,3$) are the diagonal cofactors and the three roots (say, $\lambda_{1,2,3}$) of the equation are the three eigenvalues of $J$. Evaluating the trace, deteminant and cofactors, we obtain an equation of the form
\begin{equation}
\lambda^{3}+A\lambda^{2}-B\lambda +C =0
\label{ce}
\end{equation}
where $A$, $B$ and $C$ are positive quantities given by
\begin{eqnarray*}
A &=& r_{a}\rho_a^{\ast}+r_{b}\rho_b^{\ast}+r_{c}\rho_c^{\ast} \\
B &=& \rho_a^{\ast}\rho_b^{\ast}\rho_c^{\ast}\left(\dfrac{r_{a}p_{a}}{\rho_c^{\ast}} + \dfrac{r_{b}p_{b}}{\rho_a^{\ast}} + \dfrac{r_{c}p_{c}}{\rho_b^{\ast}} \right) \\
C &=& \rho_a^{\ast}\rho_b^{\ast}\rho_c^{\ast}p_{a}p_{b}p_{c}\left( 1 + \dfrac{r_{a}}{p_{c}} + \dfrac{r_{b}}{p_{a}} + \dfrac{r_{c}}{p_{b}}\right)
\end{eqnarray*}
The three roots of Eq.~(\ref{ce}) have to abide by the following conditions
\begin{eqnarray}
\lambda_{1}+\lambda_{2}+\lambda_{3} &=& -A \label{c1}\\
\dfrac{1}{\lambda_{1}}+\dfrac{1}{\lambda_{2}}+\dfrac{1}{\lambda_{3}} &=& \dfrac{B}{C} \label{c2}
\end{eqnarray}
Note that the RHS of condition (\ref{c1}) is negative and that of condition (\ref{c2}) is positive. This indicates that all the roots (or their real part) do not have same sign. We skip the straight forward derivation of $\lambda_{1,2,3}$ as we need to know the signs of their real part only for the stability analysis. For this purpose, we utilize Descarte's rule of signs and find that Eq.~(\ref{ce}) has maximum $2$ real positive roots and $1$ real negative root. This means that, along with $1$ real negative root, Eq.~(\ref{ce}) has either $2$ real positive roots or $2$ complex roots which must be conjugate to each other. In the case of $2$ roots being complex (say, $\alpha \pm i\beta$), the real part of them must be positive, unless condition (\ref{c2}) would not be satisfied because the sum of the real part of LHS would have been negative then.

Therefore both the possibilities mentioned above eventually ensure the fact that all the eiganvalues (or their real part) are not negative for the fixed point $FP3$. Hence this point will always act as an unstable fixed point.

\end{document}